\documentclass[12pt]{article} 
\usepackage{amsfonts,epsfig} 
\usepackage[usenames]{color}  
\usepackage{times,psfrag}
\usepackage{amssymb}

\newcommand{\zsl}{\!\!\not\!{z}}
 
\newcommand{\bra}[1]{\left\langle #1 \right|} 
\newcommand{\ket}[1]{\left| #1 \right\rangle}

\newcommand{\beq}[1]{\begin{equation}\label{#1}} 
\newcommand{\eeq}{\end{equation}} 
 
\newcommand{\beqar}[1]{\begin{eqnarray}\label{#1}} 
\newcommand{\eeqar}{\end{eqnarray}}

\newcommand{\nn}{\nonumber}  
\newcommand{\dd}{{\rm d}}  

\newcommand{\ep}{\varepsilon}  
\newcommand{\ga}{\gamma}

\newcommand{\la}{\lambda}

\begin{document}  
%
%
%
\begin{titlepage}  
  
\vspace*{-2cm}  
\begin{flushright}  
\begin{tabular}{l}  
hep-ph/0311082
\end{tabular}  
\end{flushright}  
  
\vskip2.5cm  
  
\begin{center}  
{\Large \bf Improved Light-Cone Sum Rules for the Electromagnetic 
\vspace{0.5cm} \\ 
Form Factors of the Nucleon}
\vspace{1cm}  
\end{center}  
\centerline{A.~Lenz, M.~Wittmann}
\vspace{0.5 cm}  
\centerline{{\em  Institut f{\"u}r Theoretische Physik,  
Universit{\"a}t Regensburg,  
D-93040 Regensburg, Germany}}  
\vspace{1 cm}  
\centerline{E.~Stein}
\vspace{0.5 cm}  
\centerline{{\em  Physics Department,  
Maharishi University of Management,  
NL-6063 NP Vlodrop, Netherlands}}  
\vspace{1cm}  
\bigskip  
\centerline{\large \em \today}  
\bigskip  
\vfill  
\begin{center}  
  {\large\bf Abstract\\[10pt]} \parbox[t]{\textwidth}{  
We calculate the electromagnetic form factors of the nucleon within the 
light-cone sum rule approach. In comparison to previous work 
\cite{BLMS} we suggest to use a pure isospin-1/2 interpolating 
field for the nucleon, since the 
Chernyak-Zhitnitsky current leads to numerically large, unphysical, 
isospin violating contributions.
The leading-order sum rules are derived for the form factors
and the results are confronted with the experimental data. 
Our approach tends to favor the nucleon distribution amplitudes that 
are not far from the asymptotic shape.} 
  \vskip1cm  
\end{center}  
  
\vspace*{1cm}  
  
\vspace*{\fill}  
\eject  
\end{titlepage}  
%
%
%
%
%
%
%
%

\vspace*{0.2cm}

{\large\bf 1.}~~The elastic scattering of electrons off nucleons at 
momentum transfer $-Q^2$
is described by the famous Rosenbluth formula \cite{Ros50}
\beq{i.4}
    \left( \frac{d\sigma}{d\Omega} \right)  = 
          \left( \frac{d\sigma}{d\Omega} \right)_{\rm Mott}
          \left[
      \frac{ G_E^2 (Q^2) + \tau G_M^2 (Q^2)}{1 + \tau }  + 2 \tau  G_M^2 (Q^2) \tan^2 \frac{\theta}{2}
          \right],
\eeq    
where $G_E(Q^2)$ and $G_M(Q^2)$ are the electric and magnetic 
Sachs form factors,  $ \tau = Q^2/(4m^2)$, $m$ is the nucleon mass 
and $\theta$ is the scattering angle in the laboratory frame.
$\left( d\sigma/d\Omega \right)_{\rm Mott}$ is the Mott-cross section, 
which describes the scattering of a pointlike particle.
The normalization of the form factors at $Q^2=0$ is given by the nucleon charges
and magnetic moments (in units of the nuclear magneton, $\mu_N=e/2m_p$):
\beqar{i.5}
{\rm Proton:}\quad G_{E}(0)=1,&&G_M(0)=\mu_p=2.792847337(29) \; 
\cite{Hagiwara:fs}\nn\\
{\rm Neutron:}\quad G_{E}(0)=0,&&G_M(0)=\mu_n=-1.91304272(45) \; 
\cite{Hagiwara:fs} 
\eeqar
In the Breit frame 
$G_E(Q^2)$ and $G_M(Q^2)$ can be interpreted as the Fourier transforms of the 
charge distribution and magnetization density in the nucleon, respectively.
The matrix element of the electromagnetic current 
($ j_{\mu}^{\rm em}(x) = e_u \bar{u}(x) \ga_{\mu} u(x) + e_d \bar{d}(x) \ga_{\mu} d(x) $)
taken between two nucleon states is conventionally written in terms of the 
Dirac and Pauli form factors $F_1(Q^2)$ and $F_2(Q^2)$, respectively.
\beq{i.6}
\bra{P-q}j_{\mu}^{\rm em}(0)\ket{P} =
\overline{N} (P-q)\left[\ga_{\mu}F_1(Q^2)-i\frac{\sigma_{\mu\nu}q^{\nu}}{2m}F_2(Q^2)\right]N(P),
\label{formfactor}
\eeq
where $P_\mu$ is the four-momentum in the initial nucleon state, $m$ 
is the nucleon mass, $P^2= (P-q)^2=m^2$, $q_\mu$ is the (outgoing) photon momentum, 
$Q^2 = -q^2$, $\sigma_{\mu\nu}=\frac{i}{2}[\ga_{\mu},\ga_{\nu}]$  and $N(P)$ is the spinor of 
the nucleon.
The electric and magnetic Sachs form factors are related to
the Dirac and Pauli form factors in the following way
\beqar{i.7}
G_M(Q^2)= F_1 (Q^2)+             F_2 (Q^2), \; \; \; & & \; \; \;
G_E(Q^2)= F_1 (Q^2)-\frac{Q^2}{4m^2}F_2 (Q^2).
\eeqar
It is known that the experimental data for $G_M(Q^2)$ at values of  
$Q^2$ up to 5 GeV$^2$  are very well described by the famous dipole formula
both for the proton  
\cite{Hof56,Janssens66,Litt70,Berger71,Andivahis94,Walker94}
and for the neutron \cite{Lung93,Anklin98,Kubon02}
(following \cite{Arrington:2003df} we compare our theoretical predictions
only with data sets where both forward and backward angle data were taken in
the same apparatus).
\beqar{dipole}  
\frac{1}{\mu_p} G_M^{p}(Q^2)  
\sim  
\frac{1}{\mu_n} G_M^{n}(Q^2) \sim  
\frac{1}{(1+Q^2/\mu_0^2)^2}  = G_D(Q^2)\,;
\qquad \mu_0^2 \sim 0.71\,{\rm GeV^2} \; .  
\eeqar  
For the electric form factor of the proton the experimental situation  
currently is unclear. Older measurements based on the 
Rosenbluth separation showed a dipole behavior 
\cite{Janssens66,Litt70,Berger71,Andivahis94,Walker94} of
the electric Sachs form factor, but in recent measurements at the 
Jefferson Lab Hall~A Collaboration using the recoil polarization technique
a significant deviation from the dipole was observed 
\cite{Jones00,Gayou01,Gayou02}. This experimental discrepancy has been
attracting lots of attention and has not been settled yet 
(for a review see \cite{Arrington:2003df}). The values of the electric 
form factor of the neutron are very small \cite{Lung93,Zhou01,Rohe99}.
\\
The ultimate goal of the theoretical and experimental analysis of the
form factors of the nucleon is the determination of the nucleon wave 
functions. In recent years it has been becoming increasingly clear that 
the strict perturbative approach based on QCD factorization and 
involving at least two hard gluon exchanges is not applicable in the 
several GeV region and it has to be complemented by some non-perturbative
techniques. The method of light-cone sum rules (LCSR) \cite{LCSR} 
suggests itself since it incorporates both the perturbative and 
non-perturbative end-point contributions and allows to calculate the 
form factors as a systematic expansion in terms of nucleon distribution 
functions of increasing twist \cite{earlybaryon,Che84,BFMS}. 
Alternative models to determine the form factors of the nucleon can be found 
e.g. in \cite{nn}.
The general concept of the LCSR calculation is familiar from numerous 
applications of this technique to meson decays \cite{meson} and 
the particular realization for baryons
was worked out in Ref.~\cite{BLMS}. The starting point of the 
LCSR approach is that one of the participating nucleons is substituted 
by a suitable local current. The choice of the current is a subtle 
issue and is motivated by the necessity to have a strong ``nucleon signal''
and small sensitivity to the contributions of higher resonances and the 
continuum. In addition, the choice is influenced by the particular tasks 
of the calculation. In particular, in \cite{BLMS} the so-called 
Chernyak-Zhitnitsky (CZ) nucleon current was used since it allows to 
enhance contributions to the sum rule that are due to the leading twist 
distribution amplitude of interest and suppress higher-twist contributions. 
The main point of this letter is to point out that the use of the CZ current 
induces large implicit isospin violations in the sum rules of order 
20\% (and more) but this deficiency can be overcome by using a modified 
current which is a pure isospin-1/2 state. In addition to exact isospin 
symmetry, using the improved current one gets a better stability 
of the sum rules and a surprisingly 
good agreement with the experimental data using the set of 
asymptotic distribution amplitudes. We, therefore, argue that using 
the pure isospin current is advantageous and 
allows to increase the accuracy and reliability of the sum rules. 
Further applications e.g. to axial form factors will be considered 
in a subsequent publication \cite{toappear}.  
%
%
%
%
%
%
%
%

\vspace*{0.2cm}

{\large\bf 2.}~~We start with the electromagnetic coupling of protons and consider
the following correlation function  
\beqar{corr1}  
T_\nu^{\rm em}(P,q) = i \int \dd^4 x \, e^{i q \cdot x}  
\bra{0} T\left\{\eta^p(0) j_\nu^{\rm em} (x)\right\} \ket{P} 
\label{correlator}
\eeqar  
which includes an interpolating proton field $\eta^p$.
The basic principle of sum rules is to calculate this correlation function in 
two ways and finally compare the two results. First one can insert a 
complete set of states between $\eta^p$ and $ j_\nu^{\rm em}$
in Eq.~(\ref{correlator}).
\beqar{22.5}
T^{\rm em}_{\nu}(P,q)= \sum\limits_{\la,s} && 
\bra{0} \eta^p(0)\ket{\la;{ P-q},s} \frac{1}{m_\la^2- (P-q)^2}
\bra{\la;{P-q},s} j^{\rm em}_{\nu}(0) \ket{P},
\label{complete}
\eeqar
where $\lambda$ characterizes the state and $s$ stands for the polarization.
In \cite{BLMS} the CZ current \cite{Che84}  
\beqar{CZ}  
\eta^p_{\rm CZ}(0) &=& \ep^{ijk} \left[u^i(0) C \!\not\!{z} u^j(0)\right] \,  
\ga_5 \!\not\!{z} d^k(0) \,
\eeqar  
was used for $\eta^p$. In this case 
\beqar{normal}  
\label{normalization}
 \bra{0} \eta^p  \ket{P} & = & f_{\rm N}\,  
(P  z) \!\not\!{z} N(P)  
\eeqar  
(here $z$ is a light-cone vector, $z^2 = 0$), and the coupling $f_N$ 
determines the normalization of the leading twist proton distribution 
amplitude \cite{earlybaryon}. Using  
the definition of the form factors in Eq.~(\ref{formfactor}) 
the contribution of the nucleon intermediate state in the correlation 
function Eq.~(\ref{correlator}) is  readily derived to be   
\beqar{dispersion}  
z^\nu T_\nu(P,q) &=& \frac{f_N}{m^2- {P'}^2} (P' z) 
\bigg\{\bigg[ 2 F_1(Q^2) \left(P' z\right) - F_2(Q^2) (q z)\bigg]  
\!\not\!{z}\,  
\nn \\ && \hspace*{3.2cm} 
{}+ F_2(Q^2)\,\left[ (P' z) + \frac12 (q z)\right]  
\frac{\!\not\!{z} \!\not\!{q}}{m}  
\bigg\} N_p(P), 
\eeqar  
where  $ P' = P-q $.
In order to get rid of terms $\sim z_\nu$ that give subdominant contributions 
on the light-cone and to simplify the Lorentz structure we contracted 
the correlation function with $z^\nu$.
Alternatively, one can calculate the correlation function
in Eq.~(\ref{correlator}) at large 
Euclidean momenta  ${P'}^2$ and $q^2 = - Q^2$ in terms of nucleon distribution 
amplitudes. To the leading order in the strong coupling one gets expressions of the form
(cf. \cite{BLMS}) 
\begin{equation}
z^\nu T_\nu(P,q) \propto  i \int \dd^4 x \int \frac{\dd^4k}{(2 \pi)^4} e^{i (q+k)  x} \frac{kz}{k^2}  
\bra{0} \epsilon^{ijk} u_i^\alpha(a_1x) u_j^\beta (a_2x) d_k^\gamma  (a_3x)  \ket{P} C_{\alpha\beta\gamma}
\label{QCD}
\end{equation}
where $C_{\alpha\beta\gamma}$ are certain coefficients (involving Lorentz structures) and  
the real numbers $a_i$ are either one or zero. By assumption $x^2 \sim 1/(P-q)^2 \to 0$ and in 
this limit the remaining 
three-quark operator sandwiched between the proton state and the vacuum  
can be written in terms of the three-quark nucleon distribution  
amplitudes of different twist $t=3,4,5,6$, see 
\cite{earlybaryon,Che84,BFMS}.
\begin{equation}
\bra{0} \epsilon^{ijk} u_i^\alpha (a_1x) u_j^\beta  (a_2x)  d_k^\gamma  (a_3x)  \ket{P} 
= \sum \limits_i F^{(i)} X^{\alpha \beta }Y^{\gamma},
\end{equation}
where $ F^{(i)} =  V^{(i)}, A^{(i)},T^{(i)}$ are vector, axial-vector and tensor 
distribution amplitudes and $X^{\alpha \beta }$ and $Y^{\gamma}$ are Dirac structures which 
are listed in \cite{BFMS}.
Equating Eq.~(\ref{dispersion}) and the QCD calculation at a certain intermediate 
momentum $(P-q)^2\sim -1$~GeV$^2$ yields a sum rule for the form factors in terms of the 
nucleon distribution amplitudes. The matching procedure involves several technical 
steps that are common for the QCD sum rule approach in general and have the purpose of
suppressing  contributions both from higher resonances and 
the continuum, and of higher-twist operators. In particular a Borel transformation is performed,
introducing the Borel parameter $M_B$ instead of $(P-q)^2$, and the nucleon contribution 
is defined by introducing a cutoff in the spectral density at $s_0 \approx (1.5\ {\rm GeV})^2$
which is approximately the mass of the Roper resonance. 
The Borel parameter $M_B$ is chosen to be in the range $1.0-1.5$~GeV, 
see \cite{BLMS,meson}
for details. 
\\
The nucleon distribution amplitudes that provide the necessary 
non-perturbative input to the 
sum rules are usually written in terms of the conformal expansion 
\cite{BFMS,conformal}.
The so-called {\it asymptotic} distribution amplitudes correspond to taking into account 
the lowest conformal spin 
only and comparing the sum rule results with the experimental data
one may hope to get an estimate for the corrections. In Ref.~\cite{BLMS} it is shown that large 
contributions of higher conformal spins are not welcome by the data
(the fact that higher terms of the conformal expansion tend to overestimate the physical 
result is already known from the pion form factor \cite{nonlocal}), but further work is needed 
in order to make this conclusion quantitative.

The question that we address in this work is whether the accuracy of the sum 
rules can be improved by the choice of the  nucleon interpolation current. 
In particular, we look at the isospin symmetry. The CZ current (\ref{CZ})
does not have a definite isospin so that isospin relations between different nucleon
distribution amplitudes are imposed as the relations between the corresponding matrix elements. 
This current has been chosen for the sum rules in \cite{BLMS} because with this choice 
(and in contrast to, e.g. the so-called Ioffe current \cite{ioffe}) the coefficients 
$C_{\alpha\beta\gamma}$ in (\ref{QCD}) are of order one for the contributions of leading twist
distribution amplitudes and are suppressed, generically, by a power of $M_B^2$ for higher twists
(to leading order in the strong coupling). The price to pay is, however, that 
in the sum (\ref{complete}) there are contributions of both isospin-1/2 and isospin-3/2
states, e.g. the $\Delta$-resonance. It is usually believed that the isospin separation
is not important since isospin-3/2 resonances are separated from the nucleon by a relatively large 
mass gap and, therefore, sufficiently strongly suppressed by the Borel transformation. 
One may also speculate that summing over states with different isospin in fact makes the 
spectral density more smooth and thus improves the duality approximation for the continuum. 
Our starting observation is that these arguments can be checked by studying the isospin 
relations for the sum rule predictions. 
If one determines only the electromagnetic form factors of the nucleon,
as it was done in \cite{BLMS}, the necessity to fulfill isospin 
symmetry is hidden. If, however,  one determines in addition to $F_{1,2}^{pp}$
(proton in the initial and final state) and $F_{1,2}^{nn}$ (neutron in the 
initial and final state) the form factors $F_{1,2}^{np}$, which arise in the 
vector part of the weak-current ($j^{weak}_\nu (x) = \bar{u}(x) \gamma_\nu (1-\gamma_5) d(x)$)
triggering the $\beta$-decay, one can show that the isospin relation 
\begin{equation}
\label{isospin}
F_{i}^{np} = F_{i}^{pp} - F_{i}^{nn} \; \; \mbox{ for } i = 1,2
\end{equation}
has to hold. 
Checking whether Eq.~(\ref{isospin}) holds numerically for the sum rule predictions, 
we can test the assumption that the contamination by isospin-3/2 contributions in the sum rules is
negligible. The corresponding calculations (see \cite{toappear}) yield the following result:
if one uses asymptotic distribution amplitudes, then the 
isospin sum rule in Eq.~(\ref{isospin}) is violated by $\sim 20$ \%. 
If higher conformal spin contributions of the distribution amplitudes 
are taken into account, the isospin violations become even larger.
In other words, the use of the CZ current $\eta_{CZ}$ for the
evaluation of the nucleon form factors leads to an unphysical uncertainty
of at least 20 \%  induced by the ``pollution'' of sum rules by the isospin-3/2
contributions. 
\\
The problem can be overcome in a rather simple way by using a modified current 
which is a isospin-1/2 eigenstate. In particular, we suggest to use
\beq{203}
\label{inter2}
\eta^p_I(x)=\frac{2}{3} \epsilon^{ijk} \left( \left[u^i(x) C \zsl u^j(x)\right] 
\ga_5 \zsl d^k(x) -  \left[u^i(x) C \zsl d^j(x)\right]\ga_5 \zsl u^k(x)\right)
\eeq
which is an isospin-1/2 eigenstate and it projects on the leading-twist 
distribution amplitudes as well so that all ``good'' properties of the CZ current
are retained. The factor 2/3 in Eq.~(\ref{203}) is introduced to fulfill 
the same normalization condition (\ref{normalization}), so that the ``hadronic''
part of the sum rule (\ref{dispersion}) remains intact. On the other hand, 
using the improved current $\eta_I$ for the quark level calculation the 
isospin relations in Eq.~(\ref{isospin}) are recovered exactly. 
In order to be able to argue that the modified current in (\ref{203}) is indeed 
superiour for the LCSR calculations, we still need to check 
what happens with the sum rule predictions. Since in \cite{BLMS} it was found 
that large corrections to the asymptotic distribution amplitudes seem to be 
in contradiction to the data, in this letter we only consider asymptotic 
distributions as an example. A general case will be studied in (\cite{toappear}).
The final LCSRs using the improved current $\eta_I$ read
\begin{eqnarray}
F_1^{p}& = & \frac{2e_u}{3f_N} \left\{ 
\left( \int\limits_{x_1^0}^{1}\! \dd x_1 
\left[ 
\rho_1 
+\frac{m^2}{M_B^2} \left(\rho_2-\rho_3 \right)
+\frac{m^4}{M_B^4}\rho_4
\right] (x_1)  
 \;  \; \mbox{EXP}_1 \right)
\right.
\nn \\
& &
\left.
+ \left[
\left(\rho_2-\rho_3 +\frac{m^2}{M_B^2}\rho_4\right)(x_1^0)
- m^2 \frac{\dd }{\dd x_1}\ \frac{x_1^2\ \rho_4(x_1)}{Q^2+x_1^2\ m^2}
\Bigg\arrowvert_{x_1=x_1^0}\right]
\phantom{+}\frac{ m^2\ (x_1^0)^2\ }{Q^2+(x_1^0)^2\ m^2} 
\; \; \mbox{EXP}_2
\right\}
\nn\\
&&
+\frac{1}{3f_N} e_d
\Bigg\{ x_1 \to x_3, u \to d \Bigg\},
\label{F1}
\\
F_2^p&=&\frac{4e_u}{3f_N} \left\{ 
-\frac{m^2}{M_B^2}
\left( \int\limits_{x_1^0}^{1}\! \frac{\dd x_1}{x_1}
\left[ \rho_2+\frac{m^2}{M_B^2}\rho_4 \right](x_1)
\; \; \mbox{EXP}_1
\right)
\right.
\nn\\
&&
\left.
- \left[
\left(\rho_2 + \frac{m^2}{M_B^2} \rho_4\right)(x_1^0) -x_1^0 m^2 \frac{\dd }{\dd x_1}\ \frac{x_1\ \rho_4(x_1)}{Q^2+x_1^2\ m^2}\Bigg\arrowvert_{x_1=x_1^0}
\right]
\frac{m^2\ x_1^0\ }{Q^2+(x_1^0)^2\ m^2} 
\; \; \mbox{EXP}_2
\right\} 
\nn\\
&&
+\frac{2}{3f_N} e_d\Bigg\{ x_1 \to x_3, u \to d \Bigg\} ,
\label{F2}
\end{eqnarray}
where for asymptotic distribution amplitudes 
\begin{eqnarray}
\mbox{EXP}_1 & := & 
\exp\!\left( \!- \frac{1\!-\!x_1}{x_1} \frac{Q^2}{M_B^2} + x_1 \frac{m^2}{M_B^2} \right),
\nonumber \\
\mbox{EXP}_2 & := & 
\exp\!\left(\!-\frac{s_0-m^2}{M_B^2}\right),
\nonumber \\
\rho_1 (x) & = & 
60 \left( 1 - x\right)^3 x f_N,
\nonumber \\
\rho_2 (x) & = &  \frac{1}{18}
\left( 1 - x\right)^2
\left[
6 x \left(1 - 4 x\right) \lambda_1 + 
\left(36 - 370 x + 1006 x^2 - 117 x^3\right) f_N 
      \right],
\nonumber \\
\rho_3 (x)& = & - \frac{1}{72}
\left(1 - x\right)^3 x 
\left[
8 \left(9 \lambda_1 - 2 \lambda_2\right) - 3 \left(565 - 417 x\right) f_N
\right],
\nonumber \\
\rho_4 (x)& = & \frac{1}{180}
\left(1 - x\right)^3 x^2 
\left[48 \lambda_1 - 5 \left(343 - 15 x\right) f_N\right],
\nonumber \\
x_i^0 & = &  
\frac{1}{2 m} 
\left[ \sqrt{\left( Q^2 + s_0 - m^2 \right)^2 + 4 m^2 Q^2} 
- \left( Q^2 + s_0 - m^2 \right) \right].
\end{eqnarray}
The final result depends on the two ratios $\lambda_1/f_N$ and
$\lambda_2/f_N$ of the non-perturbative parameters $f_N = (5.3 \pm 0.5) \cdot 10^{-3}$ GeV$^2$,
$\lambda_1 = -(2.7 \pm 0.9) \cdot 10^{-2}$ GeV$^2$ and $\lambda_2 = (5.1 \pm 1.9) \cdot 10^{-2}$ GeV$^2$, 
which are discussed e.g. in \cite{BFMS}.
%
%
%
%
%
%
%
%
%
%
%
%
%
%

\vspace*{0.2cm}

{\large\bf 3.}~~The comparison of the sum rule results (\ref{F1}), (\ref{F2})
with the experimental data is shown in Figs.~1--5. In all cases the central 
value of the LCSR prediction is shown by the solid curve while dashed 
curves show the effect of the variation of the normalization $\lambda_1/f_N $ 
in the range $-5.1 \pm 1.7$ which is representative of the possible 
uncertainty.  Varying the Borel parameter $M_B$ in the range of $1.2$ GeV 
to $1.6$ GeV yielded no sizeable effect; in the plots  $M_B = \sqrt{2}$ GeV 
is used. 
\\
In Fig.~1 we plotted the magnetic form factor of the proton 
normalized to the dipole formula. 
In this case the difference compared to using the CZ current 
appears to be small and our results are close to \cite{BLMS}.
In both calculations the LCSR prediction using 
asymptotic distribution amplitudes tends to overestimate 
the form factor by about 50$\%$. This disagreement may signal that 
contributions of higher conformal spin have to be included, but in 
order to make quantitative 
statements one first has to calculate perturbative $\cal{O}(\alpha_s)$ corrections  
to the sum rules which is beyond the tasks of this letter.
The ratio of the electric and the magnetic proton form factors is shown 
in Fig.~2. 
Here the LCSR prediction is surprisingly close to 
the experimental values and tends to favor the values 
obtained by the recent experiments at Jefferson Lab 
\cite{Jones00,Gayou01,Gayou02}. However, in this case as well,
without the inclusion of  $\alpha_s$-corrections it is premature to draw 
definite conclusions. The difference to the calculation in \cite{BLMS} 
is quite sizeable for this ratio, up to 50\%. 
In Fig.~3 and Fig.~4 the magnetic and the electric form factors of the 
neutron are plotted, respectively. The LCSR prediction tends to 
overestimate the magnetic form factor by about 25$\%$ while for the 
electric form factor both the experiment and the LCSR give comparable 
small values. In this cases we again observe a noticeable improvement 
compared to \cite{BLMS}. 
Finally, in Fig.~5 we study the ratio $F_2/F_1$ for the proton multiplied by $Q$.
We actually plotted $Q F_2/(\kappa_p F_1)$, with the anomalous magnetic moment 
of proton $\kappa_p$, in order to have the same normalization as  the figures in 
\cite{Gayou02}.
The LCSR calculation shows a very weak dependence of this ratio on 
$Q^2$ which agrees with the scaling observed at Jefferson Lab 
\cite{Jones00,Gayou01,Gayou02}. In the LCSR approach such behavior 
results from an interplay of soft and hard contributions with different 
scale dependence and only holds approximately in a limited range of the 
momentum transfer. 
\\
To summarize, in this work we have presented arguments for the use of 
the improved nucleon current (\ref{inter2}) in the LCSR calculations. 
Our current retains all desired properties of the CZ current and in 
addition it is a proper isospin eigenstate so that isospin relations 
between form factors are valid in this case at the level of correlation 
functions and are reproduced identically in the sum rule results for the 
form factors. Our numerical estimates demonstrate that using the improved 
current one eliminates an implicit uncertainty of the calculations in 
\cite{BLMS} that is due to the isospin symmetry violation and also in  
all cases we obtain a better stability of LCSRs and a better agreement 
with the data using the set of asymptotic three-quark nucleon distribution 
amplitudes up to twist-6 constructed in \cite{BFMS}. More details and the 
application to nucleon axial form factors will be considered in a 
forthcoming publication \cite{toappear}. It has to be mentioned that the 
LCSRs to leading-order accuracy in the QCD coupling only take into account 
contributions of ``soft'' or ``end-point'' regions that are subleading 
in the true $Q^2\to \infty$ limit. The leading contributions appear at the level 
of perturbative corrections to the sum rules and their evaluation presents 
an important task for further studies. We believe that LCSRs with radiative 
corrections included can  provide quantitative information on 
nucleon distribution amplitudes.  

\vspace*{0.2cm}

{\bf Acknowledgements:}~~
We would like to thank V.M. Braun for many enlightening discussions and
reading the manuscript, N. Mahnke for useful discussions, 
M.K. Jones  for providing the data of $QF_2/F_1$ which were plotted in 
fig. 4 of \cite{Gayou02} and J. Arrington for useful comments on the 
experimental situation.
%
%
%
%
%
%

%
%
\begin{figure}[ht]
\psfrag{Y}{\large $\rm G_M^p/(\mu_P G_D)$}
\psfrag{X}{\large ${\rm Q}^2$}
\centering
  \includegraphics[width=0.75\textwidth,angle=0]{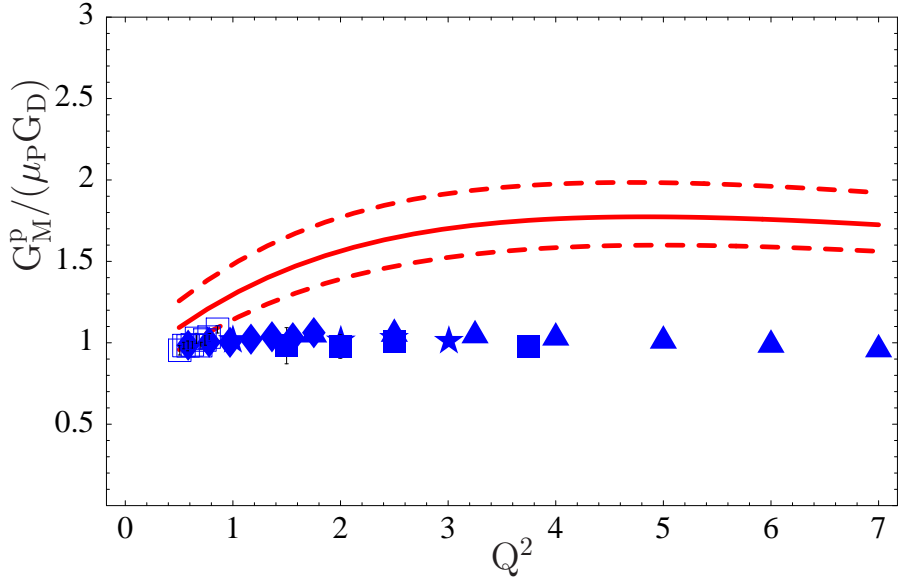}
\caption{
{\it Solid red line}: LCSR prediction for the magnetic form factor of 
the proton normalized to the dipole form factor {$\rm G_M^p/(\mu_P G_D)$}. 
{\it Dashed red lines}: errors due to the variation of the normalization 
$\lambda_1/f_N$. 
{\it Blue symbols}: experimental values: 
$\bigstar$: SLAC 1994 \cite{Walker94};
$\blacktriangle$: SLAC 1994 \cite{Andivahis94};
$\blacksquare$: SLAC 1970 \cite{Litt70}*;
$\blacklozenge$: Bonn 1971 \cite{Berger71}*;
$\Box$        : Stanford 1966 \cite{Janssens66}* 
($*$: Data actually taken from \cite{Arrington:2003df}).}
\end{figure}
%
%
\begin{figure}[b]
\psfrag{Y}{\large$\rm \mu_P G_E^p/G_M^p$}
\psfrag{X}{\large ${\rm Q}^2$}
\centering
  \includegraphics[width=0.75\textwidth,angle=0]{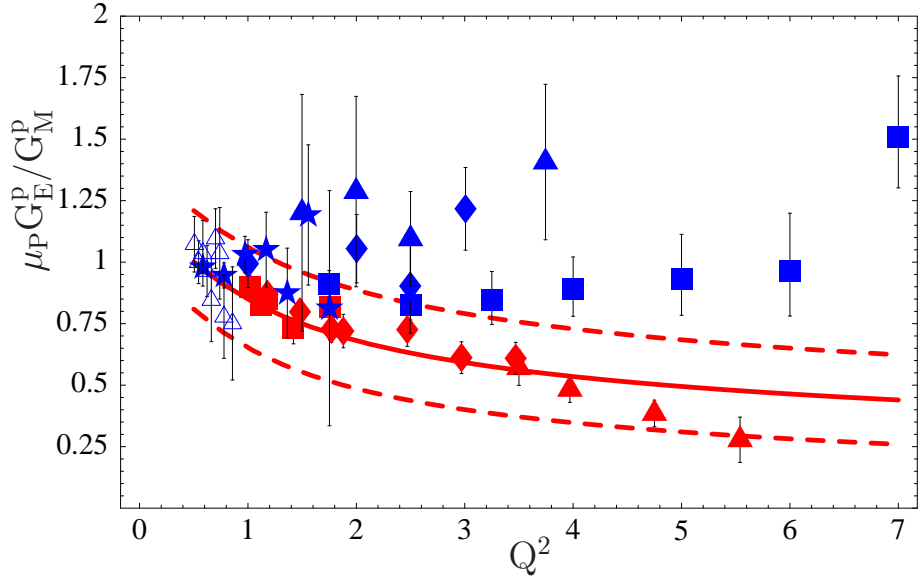}
\caption{{\it Solid red line}: LCSR prediction for the ratio of the
electric and magnetic form factors of the proton $\rm \mu_P G_E^p(Q^2)/G_M^p(Q^2)$.
{\it Dashed red lines}: errors due to the variation of the normalization 
$\lambda_1/f_N$. 
{\it Red symbols}: experimental values obtained via Polarization transfer:  
{ $\blacktriangle$: Jefferson LAB 2002 \cite{Gayou02}};
 { $\blacksquare$: Jefferson LAB 2001 \cite{Gayou01}};
 { $\blacklozenge$: Jefferson LAB 2000 \cite{Jones00}};
{\it Blue symbols}: experimental values obtained via Rosenbluth
separation:  
 { $\blacksquare$: SLAC 1994 \cite{Andivahis94}};
 { $\blacklozenge$: SLAC  1994 \cite{Walker94}}; 
 { $\blacktriangle$: SLAC 1970 \cite{Litt70}  *};
 { $\bigstar$: Bonn 1971 \cite{Berger71}*}; 
 { $\bigtriangleup$: Stanford 1966 \cite{Janssens66}*}.
($*$: Data actually taken from \cite{Arrington:2003df}).}
\end{figure}
%

\begin{figure}[ht]
\psfrag{Y}{\large$\rm G_M^n/(\mu_n G_D)$}
\psfrag{X}{\large ${\rm Q}^2$}
\centering
  \includegraphics[width=0.75\textwidth,angle=0]{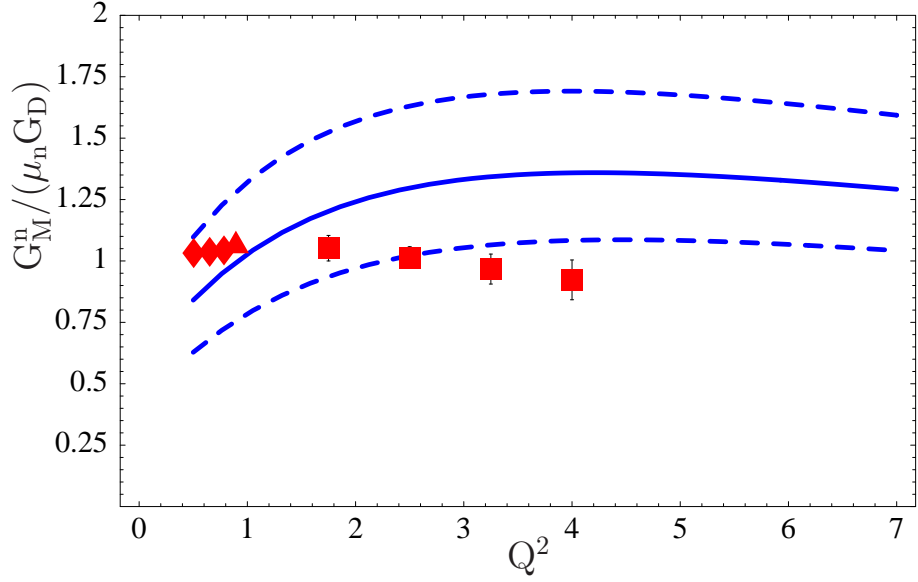}
\caption{{\it Solid blue line}: LCSR prediction for the magnetic
form factor of the neutron normalized to the dipole form factor
 $\rm G_M^n(Q^2)/(\mu_n G_D(Q^2))$.
{\it Dashed blue lines}: errors due to the 
variation of the normalization $\lambda_1/f_N$.
{\it Red symbols}: experimental values:
 { $\blacksquare$:  SLAC 1993 \cite{Lung93}};
 { $\blacktriangle$:  Mainz 2002 \cite{Kubon02}}; 
 { $\blacklozenge$:  Mainz 1998 \cite{Anklin98}}.}
\end{figure}
\begin{figure}[b]
\psfrag{Y}{\large$\rm G_E^n$}
\psfrag{X}{\large ${\rm Q}^2$}
\centering

  \includegraphics[width= 0.75\textwidth,angle=0]{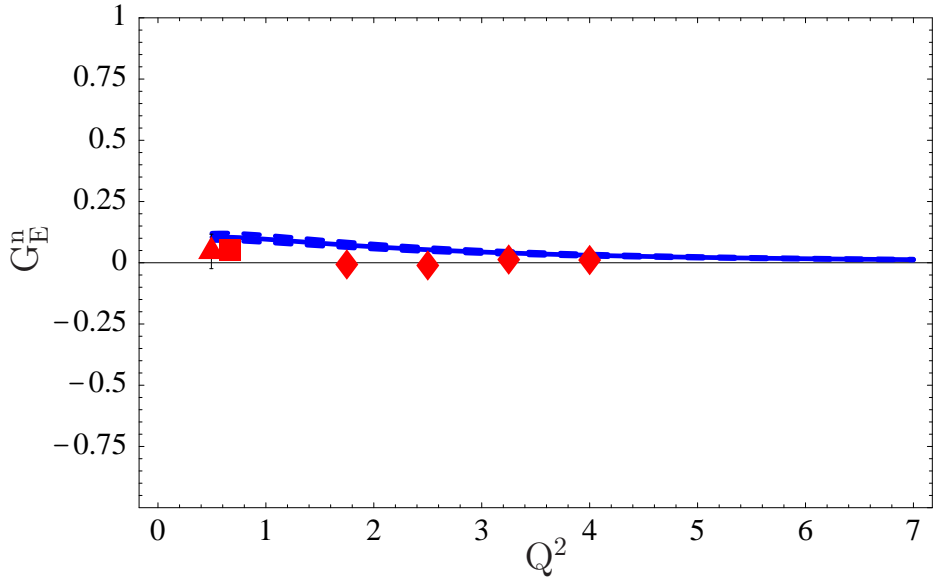}
\caption{ {\it Solid blue line}: LCSR prediction for the electric
form factor of the neutron $G_E^n(Q^2)$.
{\it Dashed blue lines}: errors due to the 
variation of the normalization $\lambda_1/f_N$.
{\it Red symbols}: experimental values:
{ $\blacklozenge$:  SLAC 1993 \cite{Lung93}};
{ $\blacktriangle$: Jefferson Lab 2001 \cite{Zhou01}};
{ $\blacksquare$: Mainz 1999 \cite{Rohe99}}.}
\end{figure}

\clearpage


\begin{figure}[ht]
\psfrag{Y}{\large$\rm Q F_2^p/(\kappa_p F_1^p)$}
\psfrag{X}{\large ${\rm Q}^2$}
\centering

  \includegraphics[width=0.75 \textwidth,angle=0]{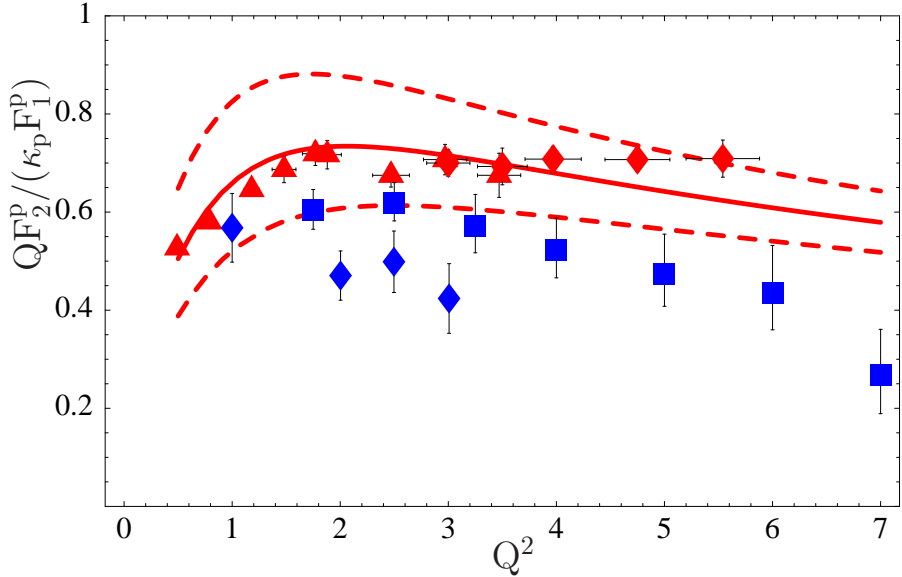}

\caption{{\it Solid red line}: LCSR prediction for the  ratio 
$Q F_2^p(Q^2)/(\kappa_p F_1^p(Q^2))$. {\it Dashed red lines}: errors due to the 
variation of the normalization $\lambda_1/f_N$.
{\it Red symbols}: experimental values obtained via Polarization transfer:  
{ $\blacktriangle,\blacklozenge$:  M. Jones (private communication)};
{\it Blue symbols}: experimental values obtained via Rosenbluth
separation:  
{ $\blacksquare$: SLAC 1994 \cite{Andivahis94}}; 
{ $\blacklozenge$: SLAC 1994 \cite{Walker94}}.}
\end{figure}

%

%
%
%
%
%

\end{document}